\documentclass{jfm}
\usepackage{graphicx}
\usepackage{epstopdf, epsfig}
\usepackage{amsmath}
\usepackage[dvipsnames]{xcolor}
\usepackage{soul}
\usepackage{siunitx}

\newcommand{\fref}[1]{figure~\ref{#1}}

\newcommand{\tref}[1]{table~\ref{#1}}
\newcommand{\sref}[1]{section~\ref{#1}}

\newcommand{\Ta}{\text{Ta}}
\newcommand{\Nuw}{\text{Nu}_\omega}

\newcommand{\aopt}{a_\text{opt}}
\newcommand{\Res}{\text{Re}_S}

\shorttitle{Double maxima of angular momentum transport in $\eta=0.91$ TC turbulence}
\shortauthor{R. Ezeta et al}

\title{Double maxima of angular momentum transport in small gap $\eta=0.91$ Taylor-Couette turbulence}

\author{
Rodrigo Ezeta\aff{1},
Francesco Sacco\aff{5},
Dennis Bakhuis\aff{1},
Sander G. Huisman\aff{1},
Rodolfo Ostilla-M\'{onico}\aff{2}
\corresp{\email{rostilla@central.uh.edu}},
Roberto Verzicco\aff{1,5,6},
Chao Sun\aff{3,1},
\and Detlef Lohse\aff{1,4}
\corresp{\email{d.lohse@utwente.nl}},
}

\affiliation{
\aff{1}Physics of Fluids Group, MESA$^{+}$ Institute  and J.M. Burgers Centre for Fluid Dynamics, University of Twente, P.O. Box 217, 7500AE Enschede, The Netherlands
\aff{2}Cullen College of Engineering, University of Houston, Houston, TX 77204, USA
\aff{3}Center for Combustion Energy and Department of Thermal Engineering, Tsinghua University, Beijing 100084, China
\aff{4}Max Planck Institute for Dynamics and Self-Organization, Am Fa{\ss}berg 17, G\"ottingen, Germany
\aff{5} Gran Sasso Science Institute, Viale Francesco Crispi 7, LAquila 67100, Italy
\aff{6} Dipartimento di Ingegneria Industriale, University of Rome “Tor Vergata”, Via del Politecnico 1, Roma 00133, Italy
}

\begin{document}

\maketitle

\begin{abstract}
We use experiments and direct numerical simulations to probe the phase-space of low-curvature Taylor--Couette (TC) flow in the vicinity of the ultimate regime. The cylinder radius ratio is fixed at $\eta=r_i/r_o=0.91$. Non-dimensional shear drivings (Taylor numbers $\Ta$) in the range $10^7\leq\Ta\leq10^{11}$ are explored for both co- and counter-rotating configurations. In the $\Ta$ range $10^8\leq\Ta\leq10^{10}$, we observe two local maxima of the angular momentum transport as a function of the cylinder rotation ratio, which can be described as either as ``co-'' and ``counter-rotating'' due to their location or as ``broad'' and ``narrow'' due to their shape. We confirm that the broad peak is accompanied by the strengthening of the large-scale structures, and that the narrow peak appears once the driving (Ta) is strong enough. As first evidenced in numerical simulations by Brauckmann \emph{et al.}~(2016), the broad peak is produced by centrifugal instabilities and that the narrow peak is a consequence of shear instabilities. We describe how the peaks change with $\Ta$ as the flow becomes more turbulent. Close to the transition to the ultimate regime when the boundary layers (BLs) become turbulent, the usual structure of counter-rotating Taylor vortex pairs breaks down and stable unpaired rolls appear locally. We attribute this state to changes in the underlying roll characteristics during the transition to the ultimate regime. Further changes in the flow structure around $\Ta\approx10^{10}$ cause the broad peak to disappear completely and the narrow peak to move. This second transition is caused when the regions inside the BLs which are locally smooth regions disappear and the whole boundary layer becomes active.
\end{abstract}


\section{Introduction}

\noindent Taylor--Couette (TC) flow, the flow in between two coaxial, independently-rotating cylinders, has successfully been used as a model for shear flows to study instabilities, flow patterns, nonlinear dynamics and transitions and turbulence \citep{GITaylor1923,Chandrasekhar1961,Andereck1986,lewis1999,pao11,vangils2011,ostilla2014b,fardin2014,Grossmann2016}. The basic TC geometry is characterized by two parameters: the first is the radius ratio $\eta=r_i/r_o$, where $r_i$ and $r_o$ are the inner and outer radii, respectively. The second is the aspect ratio $\Gamma=L/d$, where $L$ is the height of the cylinders and $d=r_o-r_i$ is the width of the gap. The shear driving of the flow is produced by the cylinders differential rotation and, in dimensionless form, is expressed by the Taylor number \citep{Eckhardt2007}

\begin{equation}
\Ta=\frac{(1+\eta)^4}{64\eta^2}\frac{d^2(r_o+r_i)^2(\omega_i-\omega_o)^2}{\nu^2},
\end{equation}

\noindent where $\omega_{i,o}$ are the inner and outer angular velocities, respectively and $\nu$ is the kinematic viscosity of the fluid. The second control parameter is the rotation ratio

\begin{equation}
    a=-\omega_o/\omega_i,
\end{equation}

\noindent where $a<0$ denotes corotation of the cylinders while $a>0$ indicates counter-rotating cylinders. The value of $a=0$ corresponds to the case of pure inner cylinder rotation. 

We note that instead of describing the control parameters of TC flow with $\Ta$, $\eta$, and $a$, one could alternatively describe the parameter space in a convective reference frame as proposed by \cite{dub05} such that the cylinders rotate with opposite velocities $\pm U/2$ and the entire system then rotates with angular velocity $\vec{\Omega} = \Omega_{rf} \vec{e_z}$ around the central axis. Here $U$ is the characteristic velocity $U=2(u_i-\eta u_o)/(1+\eta)$, $\Omega_{rf}=(r_i\omega_i +r_o\omega_o)/(r_i+r_o)$ is the mean angular velocity, $\vec{e_z}$ is the unit vector in the axial direction and $u_{i,o}=r_{i,o}\omega_{i,o}$ are the inner and outer cylinder streamwise velocities. This way any combination of differential rotations of the cylinders is parametrized as a Coriolis force. In this frame the two control parameters are the shear Reynolds number $\Res$ for the driving strength, the curvature number $R_C$, and the rotation number $R_\Omega$:

\begin{align}
    Re_S = \frac{U d}{\nu}& = 2r_ir_o \frac{d|\omega_i-\omega_o|}{(r_i+r_o)\nu}, \label{eq:Res}\\
    R_C&=\frac{(1-\eta)}{\sqrt{\eta}, }\\
    R_\Omega = \frac{2\Omega_{rf} d}{U} & =(1-\eta)\frac{r_i \omega_i +r_o \omega_o}{r_i \omega_i - r_i \omega_o}.
    \label{eq:Rom}
\end{align}

We remark that $\text{Re}_S\propto \sqrt{\Ta}$,  and that the rotation number $R_\Omega$ is connected with the negative rotation ratio $a$ by: 
\begin{equation}\label{atoromega}
    R_\Omega=(1-\eta)\frac{1-a/\eta}{1+a}.
\end{equation}

While the choice of one set of parameters might seem arbitrary at first, we note that $R_\Omega$ is the quantity that controls the magnitude of the Coriolis force when the equations are written in the rotating reference frame, and it becomes in particular relevant to elucidate certain effects, especially in the limit of low curvature \citep{hannes2016}.

In statistically stationary TC flow, the flux of angular momentum $J^\omega=r^3(\langle u_r \omega \rangle_{A,t}-\nu \partial_r \langle \omega \rangle_{A,t})$ is exactly conserved \citep{Eckhardt2007}; here $u_r$ is the radial velocity, $\omega$ the angular velocity of the fluid, $r$ is the radial coordinate and the symbol $\langle \cdot \rangle_{A,t}$ denotes a time average on a cylindrical surface coaxial with the cylinder axis. The transported quantity $J^\omega$ is independent of $r$; any flux going through an imaginary cylinder of radius $r$ also goes through any other imaginary cylinder, or mathematically $d J^\omega/d r = 0$. The response of the system can then be characterized by normalizing $J^\omega$ with its value for non-vortical laminar flow $J^\omega_{\text{lam}}=2\nu r_i^2r_o^2(\omega_i-\omega_o)/(r_o^2-r_i^2)$, which gives rise to the pseudo-Nusselt number in TC flow \citep{Eckhardt2007},
 
 \begin{equation}
    \Nuw=\frac{J^\omega}{J^\omega_{\text{lam}}}.
\end{equation}

The key scientific question is to accurately describe the transport throughout the parameter space, i.e. $\Nuw=\Nuw(\Ta,\eta,a,\Gamma)$. For low Ta, the boundary layers (BLs) remain laminar and as a consequence $\Nuw$ effectively scales roughly as $\Nuw\propto \Ta^{1/3}$ \citep{Grossmann2016}. In the ultimate regime of turbulence, in which both boundary layers and bulk are turbulent, we have $\text{Nu}_\omega\propto \Ta^{1/2} / \log\Ta$ \citep{grossmann2011}. The transition to the ultimate regime happens when the boundary layers undergo a shear instability and has been observed at $\Ta\approx \Ta_c =  3.0 \times 10^8$ for small and medium gaps ($\eta\geq0.714$) \citep{Huisman2012,ost14,Grossmann2016}. For large gaps, where curvature dominates, the transition is postponed to large values of $\Ta$, i.e.~$\Ta\approx10^{10}$ for $\eta=0.5$. If one estimates the logarithmic correction, a theoretical estimate for the effective scaling $\Nuw\propto \Ta^{0.4}$ is obtained at around $\mathcal{O}(\Ta)\approx 10^{10}$, which has been confirmed experimentally and numerically. We note that even if the value of $\Ta$ for which the transition to the ultimate regime depends on the radius ratio $\eta$ and the rotation ratio $a$, the $\Nuw(\Ta)$ effective scaling is not affected by these parameters after the transition \citep{pao11,vangils2011,merbold2013,ostilla2014b}.

While the rotation ratio does not affect the effective scaling $\Nuw\propto \Ta^{0.4}$, it has a strong effect on the proportionality constant. The rotation ratio influences the organization of the flow and increases or decreases the angular momentum transport $\Nuw$. For a fixed geometry ($\eta$), and constant driving strength (Ta), a maximum in angular momentum transport ($\Nuw$) can be found for a certain rotation ratio denoted $\aopt$ \citep{vangils2011,pao11,Grossmann2016}. For the case $\eta<0.9$, the maximum has been associated to the strengthening of the large-scale wind \citep{gil12,bra13b,Huisman2014} and the presence of turbulent intermittent bursts originated from the BLs \citep{gil12}. Beyond the point of optimal transport $a>a_{\text{opt}}$, when the counter-rotation is strong, the stabilizing effect of the outer cylinder leads to the detachment of mean vortices from the outer layer which leads to intermittent structures in the radial directions, and decreases the overall angular momentum transport \citep{bra13b}.

The value of $\aopt$ depends on the curvature of the flow, ranging from $\aopt\approx 0.2$ at $\eta=0.5$ \citep{merbold2013} to $a_{\text{opt}}\approx 0.4$ at $\eta=0.714$ \citep{Huisman2014}. \citet{ostilla2014} showed that as $\eta$ increases starting from $0.5$, $\aopt$ becomes larger, corresponding to a system with stronger counter-rotation. However, as $\eta$ increases, the peak becomes broader. To disentangle the effect of the rotation ratio $a$ from the curvature of the flow, \citet{hannes2016} numerically studied the transition from TC flow to rotating plane Couette flow (RPCF), namely the limit $\eta\to 1$ in a small-aspect ratio domain. In this limit it is more informative to look at the rotation of the cylinders as expressed by $R_\Omega$. When expressed in terms of $R_\Omega$, the asymptotic value (for $\eta\to 1$) of $R_{\Omega,\text{opt}}$ remains approximately constant. On the other hand in the limit $\eta\to 1$, $a(\eta)$ converges to $a = 1$ in a rotationless system, and to $a = -1$ for all the other cases, showing that for this parameter the transition between TC flow and RPCF is singular (see \cite{hannes2016} for further details). Strikingly, \citet{hannes2016} find that for $\eta>0.9$ (low curvature), not one maximum of angular momentum transport is present, but two. The first peak, located in the corotating regime, was described as the broad peak. It is associated with strong vortical motions, as evidenced by the radial velocity fluctuations which show a maximum at optimal transport \citep{hannes2016}. The second peak, denoted as the narrow peak, was found for counter-rotating cylinders. It appeared only when the driving is sufficiently large, and it was speculated that it supersedes the broad peak for sufficiently large driving.

The appearance of two peaks for small gaps means that several competing mechanisms for the formation of the optimum momentum transport must exist, and that these become blurred for large gaps as the stabilizing effects due to curvature add a third factor. By analyzing the $\Nuw(a)$ relationship using $R_\Omega$ as a control parameter, \citet{hannes2016} were able to show that the peaks appearing in the counter-rotating regime at $\eta=0.5$ and $\eta=0.714$, and the broad peak for corotating cylinders for $\eta>0.8$ basically were the same phenomena, as they both contained strong ordered motions and fell in the same $R_\Omega$ range. As this peak survives the limit of vanishing curvature, it becomes clear that intermittency originated from the stabilizing effect of the outer cylinder does not explain its origin. Instead, \citet{hannes2017} divided the TC system into three sub-systems in the spirit of \citet{mal54}: the bulk, and the two boundary layers representing marginally stable TC systems. With this simple model, they were able to predict the location of the broad peak in $R_\Omega$ space, finding good agreement for the prediction at moderate $\Ta$. Using the same argument, \citet{hannes2017} also predicted that the shear in the boundary layers, and hence their transition to turbulence, depends not only on the absolute shear driving, but also on the rotation ratio, which was corroborated by experiments. In this way, they explained the appearance of the narrow peak  as an enhancement of angular momentum transport in certain regions of parameter space caused by the ``early'' transition of the BLs to turbulence. \citet{hannes2017} also argued that the narrow peak will dominate the broad peak once the centrifugal instabilities are superseded by shear instabilities, and only one peak would be visible as in \cite{ostilla2014}. This was postulated to happen once the BLs become turbulent for the value of $a$ in the broad peak. \cite{hannes2017} predicted this to happen at $\Ta>4.95\times 10^{9}$, close to the transition to the ultimate regime for that $\eta$.

In this study we set out to globally and locally probe the angular momentum transport in a wide range of driving strength $10^7\leq\Ta\leq10^{11}$ for the case of low curvature $\eta=0.91$, focusing in particular in the $\Ta$ range $10^8\leq\Ta\leq10^{10}$, where the transition to the ultimate regime happens, and where \citet{hannes2017} observed the appearance of two peaks for angular momentum transport using numerical simulations. The main motivation of this study is to elucidate the link between the change in behaviour of the $\Nuw(\Gamma)$ dependence \citep{mar14}, the vanishing of the broad peak, and the changing role of vortical motions \citep{sac18} which all happen around this transition. We will use an experimental setup with very large aspect ratios $\Gamma$, which allows for the flow to switch between states, i.e. different roll wavelengths. By doing this, not only can we experimentally confirm the appearance of multiple angular momentum optima in TC flow, which has not been reported yet, but we can also study the transition between regimes dominated by narrow and broad peaks. We will also rule out that they are an effect of artificially constraining the flow to small periodic aspect ratios: switching between two and three-roll states for varying driving was already reported in \cite{ostilla2014}, and this could have an effect on the two peaks.

Secondly, we will test the predictions of \citet{hannes2016} and \citet{hannes2017} regarding the mechanism underlying the occurrence of both peaks. Is the broad peak related to vortical motions, which are strengthened by centrifugal forces? Is the narrow peak a consequence of shear? And if so, will it overtake the broad peak and if so, at what turbulence level? By carefully examining the regime where the boundary layers transition, we can better explore the mixed dynamics arising when centrifugal effects and shear are competing side-by-side and further understand what is happening at the transition to the ultimate regime. To address these questions we conducted both experiments (torque and local velocity measurements) and Direct Numerical Simulations (DNS). 

The structure of the paper is as follows. In \sref{sec:exp_setup}, we explain the experimental methods. In \sref{sec:num_details}, we introduce the numerical details of the simulations. In \sref{sec:global}, we experimentally study the global response of the flow throughout a large parameter space of $\Ta$ and $a$. In particular, we reveal transitions and local maxima of the angular momentum transport. In \sref{sec:simulations}, we complement the experimental findings with numerical simulations and  discuss in detail how the size and shape of the Taylor rolls changes with varying the rotation parameter $R_\Omega$. The final \sref{sec:conclusion} contains the conclusions and an outlook for future works.

\section{Experimental setup and measurement procedure}
\label{sec:exp_setup}
\subsection{Setup}
\noindent The experiments were carried out in the Twente Turbulent Taylor-Couette facility ($\text{T}^3\text{C}$) \citep{vangils2011}. In this apparatus, the ratio $\eta$ and aspect ratio $\Gamma$ can be adjusted by installing outer cylinders of different dimensions. In this study, the radius of the inner cylinder (IC) is $r_i=\SI{200}{\mm}$ and the radius of the outer cylinder (OC) is set to $r_o=\SI{220}{\mm}$. As a consequence, the radius ratio is $\eta=r_i/r_o \approx 0.91$ and the aspect ratio results in $\Gamma=L/d=46.35$, with $d=r_o-r_i=\SI{20}{\mm}$ and $L=\SI{927}{\mm}$. Two acrylic windows located at the bottom cylinder, which cover the entire gap, allow for the capture of Particle Image Velocimetry (PIV) fields in the $r-\theta$ plane. The advantage of having a second window in the bottom plate is that we can capture two velocity fields for every revolution of the outer cylinder (see \fref{fig:figure1}).

\subsection{Global measurements: Torque}
\label{subsec:torque_measurements}
We measure the torque $\mathcal{T}$ required to drive the cylinders at constant speed. This is done via a Honeywell model 2404 hollow reaction torque sensor which connects  the  driving  shaft  and  the  inner  cylinder. The accuracy of the sensor is $0.2$ Nm. From the torque measurements, the Nusselt number can be calculated as follows

\begin{equation}
    \Nuw=\left(\frac{r_o^2-r_i^2}{2 \nu r_i^2 r_o^2 \Delta \omega} \right) \left(\frac{\mathcal{T}}{2\pi \ell_{\text{eff}} \rho } \right), 
\end{equation}

\noindent where $\ell_{\text{eff}}=\SI{536}{\mm}$ is the effective length along the cylinder where the torque is measured, the difference of angular frequencies is $\Delta \omega=2\pi \Delta f=2\pi(f_i-f_o)$ with $f_{i,o}$ the driving frequency of the inner and outer cylinder, respectively, and $\rho$ the fluid density.
Typically, the $\text{T}^3\text{C}$ facility operates in the ultimate regime of turbulence, where both boundary layers (inner and outer) are turbulent; in our case, where $\eta=0.91$, this corresponds to a driving of $\Ta>\mathcal{O}(10^8)$. Thus, in order to capture the transitional regime ($\mathcal{O}(10^7)<\textrm{Ta}<\mathcal{O}(10^8)$), we use working fluids with different values of the kinematic viscosity $\nu$. The working fluid---depending on the desired range of Ta to be resolved---is a mixture of water and pure glycerol. The percentage of glycerol in the mixtures, along with its corresponding kinematic viscosity and density  can be found in \tref{tab:table1}.  The liquid temperature is kept constant at $\SI{21}{\celsius}$ during all the experiments. 

We probe the phase space of $\text{Nu}_\omega$ in two different ways. The first one is what we call an $a$-sweep, where the angular velocity difference $\Delta \omega$ and thus the driving strength $\Ta$ is kept constant and the angular velocity ratio $a=-\omega_o/\omega_i$ is varied. In this way, we can measure different states in the co- and counter-rotation regime while the driving ($\Ta$) is fixed. The second type of experiments is the opposite i.e. a $\Ta$-sweep, where $a$ is fixed and $\Delta \omega$, and thus the driving strength $\Ta$, is increased. 

\begin{table}
    \centering
    \begin{tabular}{c|c|c|c}
       Working Fluid  & (\%) Glycerol & $\nu / \nu_w$ & $\rho / \rho_w$ \\
          \hline
         Mixture 1 & 58.72 & 18.20 & 1.18  \\
         Mixture 2 & 55.60 & 13.11 & 1.17  \\
         Mixture 3 & 45.39 & 5.67 & 1.13  \\
         Mixture 4 & 40.18 & 4.09 & 1.11  \\
         Water  & 0 & 1.0 & 1.0  \\

    \end{tabular}
    \caption{Properties of the different mixtures used in the experiments. The percentage of glycerol is based on volume. Both the density and kinematic viscosity ratios are calculated with respect to the density $\rho_w$ and the kinematic viscosity $\nu_w $ of water at $\SI{21}{\celsius}$. Data taken from \citet{cheng2008}. }
    \label{tab:table1}
\end{table}

\subsection{Local measurements: PIV}
\label{subsec:piv_measurements}

\begin{figure}
\begin{center}
\includegraphics[scale=0.25]{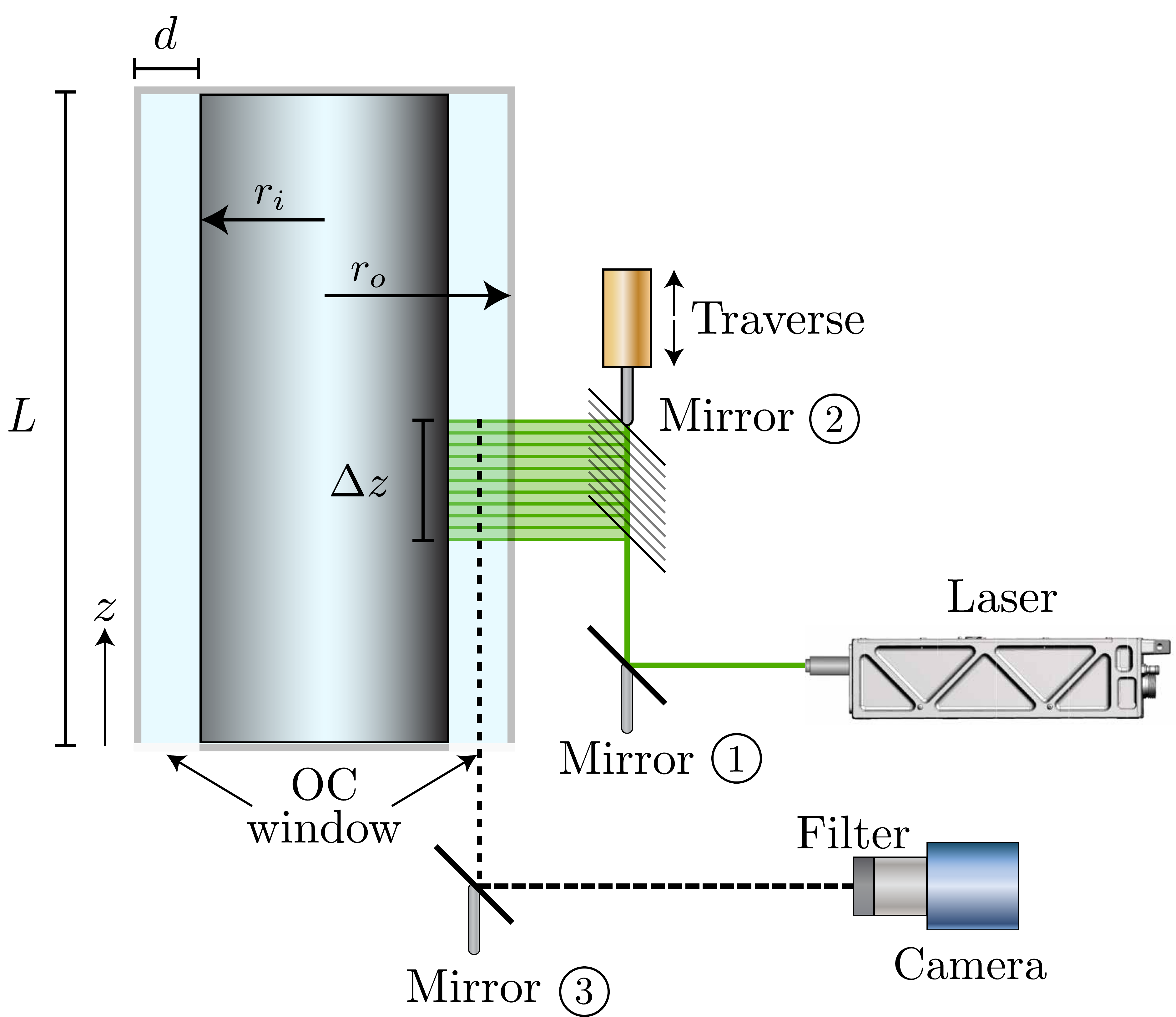}
\caption{Sketch of the experimental set-up. All elements (except for the laser) are mounted on to the frame of the $\text{T}^3\text{C}$. This results in no relative motion between the camera and the laser sheet due to mechanical vibrations. The velocity fields obtained with PIV are measured on the $r-\theta$ plane.}  
\label{fig:figure1}
\end{center}
\end{figure}

We seed the flow with polyamide fluorescent particles with diameters up to $\approx \SI{20}{\um}$ with a seeding density of $\approx 0.01 \  \text{particles} / \text{pixel}$. The emission peak of these particles is centered at $\approx \SI{565}{\nm}$. We image the particles in the flow with an \texttt{Imager SCMOS ($2560\times2160\;\text{pixel})$ 16 bit} camera using a \texttt{Carl Zeiss Milvus 2.0/100} objective. The illumination of the particles is provided by a \texttt{Quantel Evergreen 145, 532 nm} dual cavity pulsed laser. A cylindrical lens is positioned at the laser output to create a thin light sheet of $\approx \SI{1}{\mm}$ thickness. A set of mirrors and a traverse system are installed which allows the laser sheet to move with the frame of the T$^3$C  (see \fref{fig:figure1}). Explicitly, the laser beam (the laser is not mounted on to the frame) hits mirror 1 (see \fref{fig:figure1} for the labeling of the mirrors)  which is tilted $\ang{45}$. Light will then be redirected upwards towards to mirror 2 (also tilted at $\ang{45}$) which redirects it finally towards the OC, perpendicular to both cylinders. A third mirror (mirror 3) is attached to the traverse system of the $\text{T}^3\text{C}$  which can move freely in the axial direction. All elements except for the laser head, are mounted on to the frame of the $\text{T}^3\text{C}$. This results in no relative motion between the camera and the laser sheet due to mechanical vibrations while the system is rotating.

The experiments require the OC to move freely; thus, a special trigger for the camera is used for the acquisition of the images. This triggering is done by magnets located on top of the OC and a Hall switch mounted onto the frame of the T$^3$C which outputs a voltage signal every time the magnets pass by. Using this signal as a trigger, we are able to capture two fields (each one corresponding to one window in the bottom plate) per revolution of the OC. The camera is operated in double frame mode with a framerate $f$ that depends on the rotation rates of the outer cylinder $f_o$. In all cases however $\Delta t \leq 1/f$, where $\Delta t$ is the interframe time. In order to increase the contrast between the emission of the light from the particles and the background, we use an \texttt{Edmund High-Performance Longpass filter $\SI{550}{\nm}$} in front of the camera lens.

A total of 7 different flow states have been investigated using particle image velocimetry. These 7 flow states have different $a$ and $\Ta$, as reported in \tref{tab:table2} and---as will be shown later---correspond to the local maxima of the angular momentum transport as function of $a$ for a variety of $\Ta$. In total, 10 different heights were explored for each state, and 500 fields were recorded for each height. The heights are uniformly spaced with a separation of $\delta_z = \SI{10}{\mm}$, and span the length $\Delta_z = \SI{100}{\mm}$ along the axial direction. When normalized with the height of the cylinders $L$, this corresponds to a an axial span of $\Delta_z/L = (10\delta_z)/L = 0.108$ within the range $z/L\in[0.403,0.5]$. for all the experiments. The movement of the laser sheet in the axial direction results in defocusing of the images, therefore the focus is adjusted accordingly for all the explored heights. Accordingly, every velocity field is independently calibrated depending on which window of the bottom plate of the OC is used to acquire it, the height of the laser sheet, the rotation ratio and the driving. Therefore, for a fixed window, $a$ and $\Ta$, the resolution of the PIV fields depends on how far the field is seen by the camera, i.e. the height along the cylinder (see \fref{fig:figure1}). In the axial range explored in this study, the resolution of the cameras is within $\approx [30,35] \ \mu \text{m} / \text{px}$, with the lower/upper bound corresponding to the closest/furthest height from the camera respectively.

\begin{table}
    \centering
    \begin{tabular}{c|c|c|c|c}
    \hline
    \multicolumn{5}{c}{Experiments}\\
    \hline
    Ta  & $Re_s$ & $a$ & $R_\Omega$ & $\Gamma$ \\
    \hline
    $6.35\times 10^7$ & $7.93\times 10^3$ & $-0.58 \quad $ -- $\quad 0.80$  & $0.351 \quad $ -- $\quad \phantom{-}0.006 $  & $46.35$\\
    $1.33\times 10^8$ & $1.15\times 10^4$ & $-0.60 \quad $ -- $\quad 1.50$ & $0.373 \quad $ -- $\quad -0.023$  & $46.35$\\
    $1.63\times 10^8$ & $1.27\times 10^4$ &$-0.26 \quad $ -- $\quad 0.80$ & $0.156 \quad $ -- $\quad \phantom{-}0.006 $  & $46.35$\\
    $3.29\times 10^8$ & $1.81\times 10^4$ & $-0.30 \quad $ -- $\quad 0.80$ & $0.171 \quad $ -- $\quad \phantom{-}0.006 $  & $46.35$\\
    $5.10\times 10^8$ & $2.25\times 10^4$ & $-0.49 \quad $ -- $\quad 0.80$ & $0.362 \quad $ -- $\quad \phantom{-}0.006 $  & $46.35$\\
    $2.23\times 10^9$ &$4.70\times 10^4$ & $-0.33 \quad $ -- $\quad 0.80$ & $0.183 \quad $ -- $\quad \phantom{-}0.006 $  & $46.35$\\
    $3.31\times 10^9$ & $5.78\times 10^4$ & $-0.29 \quad $ -- $\quad 0.79$ & $0.167 \quad $ -- $\quad \phantom{-}0.007$  & $46.35$\\
    $1.40\times 10^{10}$ & $1.18\times 10^5$ & $-0.60 \quad $ -- $\quad 1.50$ & $0.373 \quad $ -- $\quad -0.230 $  & $46.35$\\
    $4.30\times 10^{10}$ & $2.06\times 10^5$ & $-0.30 \quad $ -- $\quad 1.00$ &$0.171 \quad $ -- $\quad -0.004 $  & $46.35$\\
    \hline
    \multicolumn{5}{c}{DNS}\\
    \hline
    Ta  & $Re_s$ & $a$ & $R_\Omega$ & $\Gamma$ \\
    \hline
    $5.10\times 10^8$ & $2.25 \times 10^4$ & $-0.761 \quad $ -- $\quad 0.909$ & $0.7 \quad $ -- $\quad 0$ & $12.56$ \\
    $1.17\times 10^9$ & $3.4 \times 10^4$ & $-0.761 \quad $  -- $\quad 0.909$ & $0.7 \quad $ -- $\quad 0$ & $2.33$  \\
    \hline
    \end{tabular}
    \caption{Experimental and numerical flow parameters used in this study. The first two columns show the driving, expressed as either $\Ta$ or $Re_S$. The third and fourth columns show the rotation parameters expressed as either $a$ or $R_\Omega$. The last column shows the aspect ratio $\Gamma$.}
    \label{tab:table2}
\end{table}

The velocity fields are measured in the $r-\theta$ plane and are computed by a multi-pass algorithm using commercial software ($\texttt{Davis 8.0}$). The first pass is set to  $64\times 64 \ \textrm{pixels}$ and the last one is set to $24\times 24 \ \textrm{pixels}$ with $50\%$ overlap of the windows. The fields obtained are then expressed in cylindrical coordinates of the form $\vec{u}=u_r(r,\theta,t)\vec{e_r}+u_\theta(r,\theta,t)\vec{e_\theta}$, where $u_r$ and $u_\theta$ are the radial and azimuthal velocities, respectively, which depend on the radius $r$, the angular (streamwise) direction $\theta$ and time $t$. Here, $\vec{e_r}$ and $\vec{e_\theta}$ are the unit vectors in the radial and azimuthal direction, respectively.

\section{Setup of the direct numerical simulations}
\label{sec:num_details}

\noindent In addition to experiments, we perform direct numerical simulation (DNS) using an energy-conserving second-order centered finite-difference code for the spatial discretization, while a fractional time-step advancement is adopted in combination with a low-storage third-order Runge-Kutta method. The complete description of the algorithm can be found in \cite{ver96} and \cite{poe15}. This code has been extensively used and validated for TC flows \citep{ostilla2014b}. 

As mentioned in the introduction, we perform the simulations in a convective reference frame \citep{dub05}, determined by the parameters $Re_s$ and $R_\Omega$ defined in equations \eqref{eq:Res} and \eqref{eq:Rom}. According to this scaling the nondimensional incompressible Navier-Stokes equations read: 

\begin{equation}
    \nabla \cdot \textbf{u}=0,
\end{equation}

\begin{equation}
 \displaystyle\frac{\partial \textbf{u}}{\partial t} + \textbf{u}\cdot \nabla \textbf{u} + R_\Omega \textbf{e}_z\times\textbf{u} = -\nabla p + Re_S^{-1} \nabla^2 \textbf{u}.
\end{equation}

We chose the same radius ratio $\eta=0.91$ as in experiments, which is also the same as in the numerical simulations of \citet{ostilla2014,ostilla2014b}. We perform two sets of simulations with fixed Reynolds numbers, $\Res =2.25 \times 10^4$ and $\Res =3.4 \times 10^4$ (or $\Ta=5.10\times 10^8$ and $\Ta=1.17 \times 10^9$) while varying $R_\Omega$ (or equivalently $a$). Axially periodic boundary conditions are taken with a periodicity length which is similar to the height of the cylinder $L$, because even if the boundary conditions are different, the resulting rolls end up being approximately the same size. Non-dimensionally this is expressed by the aspect ratio $\Gamma$. In the azimuthal direction, the system is naturally periodic; however, an imposed artificial rotational symmetry of order $n_{sym}$ is chosen in order to reduce the computational costs.

We take two computational box-sizes. A small box similar to the one used by \cite{bra13} with $\Gamma=2.33$ and $n_{sym}=20$. This small box is used for both values of $\Res$, and is large enough to not affect the first-order statistics of the flow \citep{ost15}. For the case of $\Res =2.25 \times 10^4$, we also run a medium-sized box with an aspect ratio of $\Gamma=12.56$, and a rotational symmetry of $n_{sym}=3$. This allows the flow some freedom to switch between different roll states, as in \cite{ost16vlb}. A uniform discretization is used in the azimuthal and axial directions, while a Chebychev-type clustering near the cylinders is used in the radial direction. The spatial resolution for the small boxes at $\Res=2.25\times 10^4$ and $\Res = 3.4\times 10^4$ was chosen as $n_\theta\times n_r\times n_z = 384 \times 512 \times 768$ in the azimuthal, radial and axial directions, which in wall units for the more restrictive case of $\Res=3.4 \times 10^4$ is a resolution of $\Delta z^+\approx 5$, $\Delta x^+ = r\Delta \theta^+\approx 9$ and $0.5 \leq \Delta r^+\leq 5$. For the medium-size box at $\Res =2.25 \times 10^4$, a grid of $n_\theta\times n_r\times n_z = 1728 \times 384 \times 1728$ was chosen, which yields a resolution of $\Delta z^+\approx 5$, $\Delta x^+ = r\Delta \theta^+\approx 9$ and $0.4 \leq \Delta r^+\leq 2.5$. In order to achieve temporal convergence, the simulations are run until the difference between the time-averaged torque of the inner and the outer cylinders is less than 1\%. The torque is then taken as the average between these two values. The simulations are then run for at least 40 large eddy turnover times $tU/d$.

\section{Transitions and local maxima in $\Nuw(\Ta,a)$}
\label{sec:global}

\subsection{Transitions in the $\Nuw(\Ta)$ scaling}

\begin{figure}
\begin{center}
\includegraphics[scale=0.4]{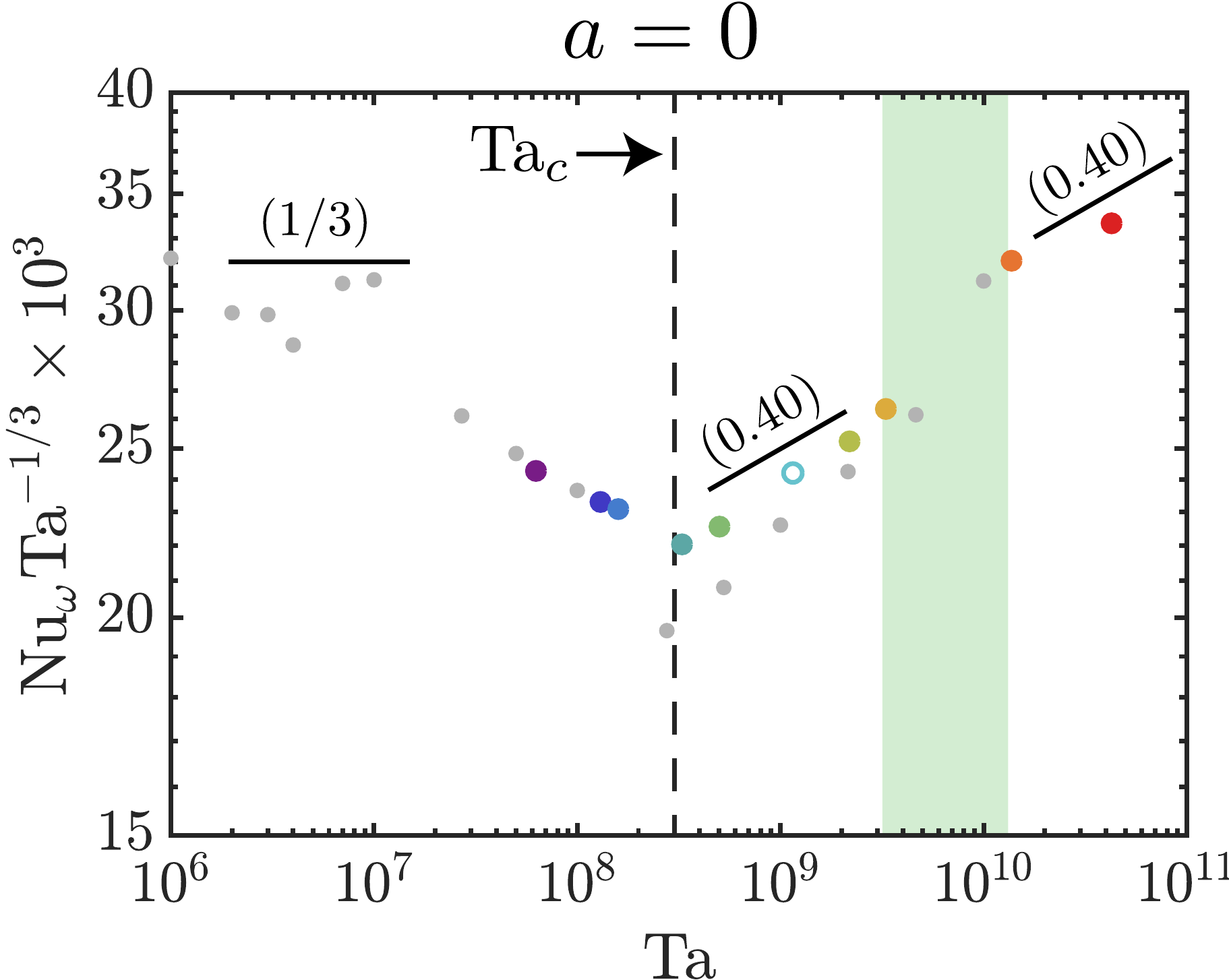}
\caption{Compensated Nusselt number as a function of the driving strength $\Ta$ for the case of pure inner cylinder rotation $a=0$ at $\eta=0.91$. The gray data points ($\text{Ta}<10^8$) correspond to DNS from \citep{ostilla2014}. The gray data points for $\text{Ta}>10^8$ are also DNS simulations but from a different study \citep{ostilla2014b}. In addition, each colored marker at fixed $\Ta$ corresponds to the driving variation as shown in the legend of \fref{fig:piv_points}c. The open circle in light blue corresponds to the DNS data of the current study for $\Gamma=2.33$. The transition to the ultimate regime is observed at $\Ta = \text{Ta}_\text{c} \approx 3 \times 10^8$ (vertical dashed line). The green shaded area corresponds to the region where a local change of slope can be seen due to the disappearance of quiescent wind shearing regions. The black solid lines serve as a reference to indicate the corresponding scaling.}
\label{fig:a_0}
\end{center}
\end{figure}

First, we analyze the scaling laws of the $\Nuw(\Ta)$ curve for pure inner cylinder rotation in \fref{fig:a_0}. The Nusselt number is compensated by the scaling of the classical regime, i.e. $\Nuw \Ta^{-1/3}$ and plotted as a function of the driving strength $\Ta$ for pure inner cylinder rotation only ($a=0$). We also include the DNS of \citet{ostilla2014b}, and observe a good agreement between the numerics and the experiments. For values of the driving $\Ta<10^7$, the flow is still in the classical regime, where both BLs are still laminar, and an effective scaling of $\Nuw \propto \Ta^{0.3}$ can be observed. When the driving strength is increased beyond $\Ta=\mathcal{O}(10^7)$, the flow enters a transitional regime, with an effective scaling exponent $\alpha$ in $\Nuw \propto \Ta^\alpha$ of $\alpha \approx 0.2$. If the driving is further increased, a minimum value of the compensated Nusselt number is reached at a critical Taylor number $\Ta_c \approx 3.0 \times 10^8$, after which a clear change in the scaling exponent to $\alpha=0.4$ can be seen. This indicates the onset of the ultimate regime, which coincides for experiments and numerics. 

Figure \ref{fig:a_0} reveals also a second phenomenon, which was not previously reported in experiments. In \cite{ostilla2014}, the local scaling-law was found to be $\Nuw\sim\Ta^{0.4}$ for $\Ta>10^{10}$. Indeed, provided $\Ta > \Ta_c$, the local effective scaling-law appears to be the same, with one caveat: there appears to be a local change of slope in the curve around $\Ta\approx 10^{10}$, where the local scaling exponent increases for a small range in $\Ta$. The region where this occurs is highlighted in green in \fref{fig:a_0}. \cite{ostilla2014b} observed a similar sudden increase in the local scaling exponent in DNS simulations at $a=0$, and attributed it to the sudden disappearance of quiescent wind shearing regions in the boundary layer. After this increase, the dependence of $\Nuw$ on the roll wavelength was completely lost \citep{ost15}. Here, we find evidence that experiments see a similar, sharp increase as observed in figure \ref{fig:a_0}. This will be further investigated in \sref{sec:merging}, where we will explore how this behaviour is seen across the $a$-range, and its effect on the local maxima of angular momentum transport.

\subsection{Appearance and shifting of the local maxima}
\begin{figure}
 \begin{center}
 \includegraphics[scale=0.3]{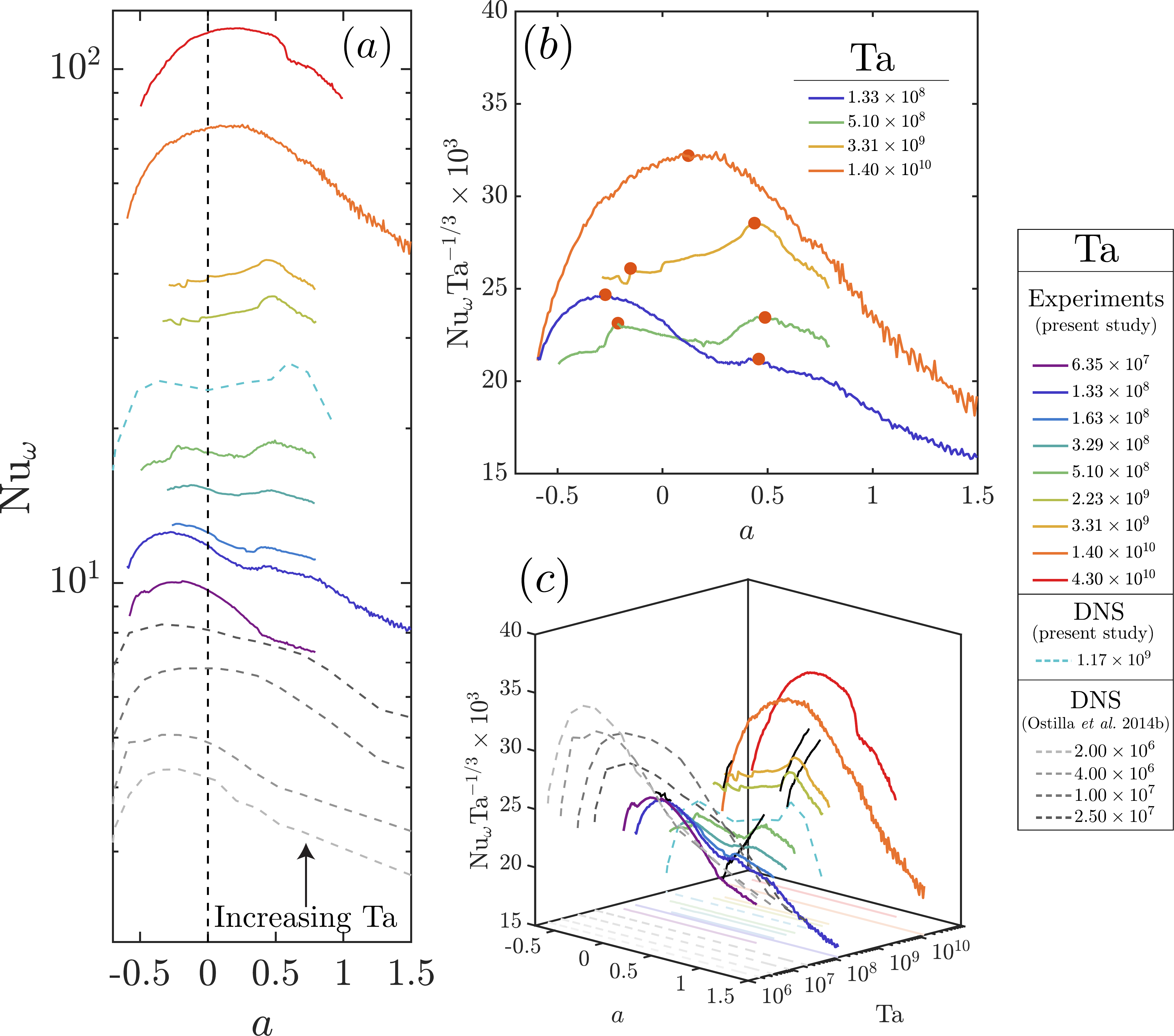}
 \caption{(a) Nusselt number $\Nuw$ as a function of rotation ratio $a$ for different values of the driving $\Ta$. (b) Compensated Nusselt number as a function of $a$ for four selected $\Ta$. Here, the solid red circles represent local maxima of angular momentum, where we perform PIV measurements as described in \sref{sec:piv}. The vertical dashed line ($a=0$) in both (a) and (b) separates the co and counter-rotating regimes. (c) A 3D representation of the compensated Nusselt number as a function of $\Ta$ and $a$. The black solid lines represent the experiments performed for fixed $a$, i.e. $\Ta$-sweeps. An animated version of this figure can be found in the Supplementary Material. In all figures, the solid lines represent experiments while the dashed lines represent numerics. The colors represent the variation in $\Ta$ as illustrated by the legend.}
 \label{fig:piv_points}
 \end{center}
 \end{figure}

Once we allow the outer cylinder to rotate, we have a more complicated three-dimensional parameter space. In \fref{fig:piv_points}a, we show the Nusselt number as a function of the rotation ratio $a$ for different $\Ta$. This figure reveals---just as the DNS from \citet{ostilla2014} and \citet{hannes2016}--- that a very pronounced maximum of angular momentum can be found in the corotating regime when the driving is $\Ta<1.33\times 10^8$. As the driving exceeds the critical value $\Ta_c$, for about a decade of $\Ta$ we can temporally identify two local angular momentum maxima: the first is located in the corotating regime at $a\approx -0.27$ and the second in the counter-rotating regime at $a\approx 0.46$. These turbulent states correspond to $R_\Omega=0.16$ and $R_\Omega=0.03$, which are similar to the values found by \citet{hannes2016} for $\eta=0.91$.  The two maxima are prominent for a very small $\Ta$ range: $3.29\times10^8<\Ta<1.17\times10^9$.

This measurement reveals that the two local angular velocity transport maxima for the same driving are not an artifact of the initial conditions or the finite extent of the domain of the numerics. As we further increase the driving beyond $\Ta>3.31\times 10^9$, the maximum in the corotating regime (broad peak) vanishes, while the maximum in the counter-rotating regime (narrow peak) increases its magnitude. For the largest driving we explore ($\Ta=4.30\times 10^{10}$), only one peak can be detected in the counter-rotating regime, although it is now less sharp. In order to highlight this trend, we show in \fref{fig:piv_points}b the compensated Nusselt number for four selected values of $\Ta$. Again, note how the value of the driving dictates the occurrence of the maximum of angular momentum transport: if $\Ta$ is too small, only one peak can be found in the corotating regime. Conversely, if $\Ta$ is too large, only one peak can be observed albeit for counter-rotation. There is, however, a range of $\Ta$ which lies in between these two extremes, for which two maxima can be detected. In \fref{fig:piv_points}c, we show a 3D representation of the compensated Nusselt number as a function of $a$ and $\Ta$. In this figure, we included the experiments for fixed $a$ (shown in black), i.e. $\Ta$-sweeps. Note how these experiments agree remarkably well with both the $a$-sweeps (shown in color) and the numerics, mutually validating each other. An animated version of this figure can be found in the Supplementary Material. We finally note that for $\Ta=2.23\times 10^9$ and $\Ta=3.31\times 10^9$, discrete jumps in the $\Nuw(a)$ can be observed for $a<0$. This observation can better be seen in \fref{fig:piv_points}a and will be revisited in $\S$\ref{sec:dns} with the results from the numerical simulations. We note however that, although these jumps are similar in magnitude to noise in other curves at different $\Ta$, we are confident that they are physical and not an artifact of the measurement system. This is based on the high accuracy of the torque sensor in these cases relative to the absolute value of the measured torque in Nm.

\begin{figure}
 \begin{center}
 \includegraphics[width=\textwidth]{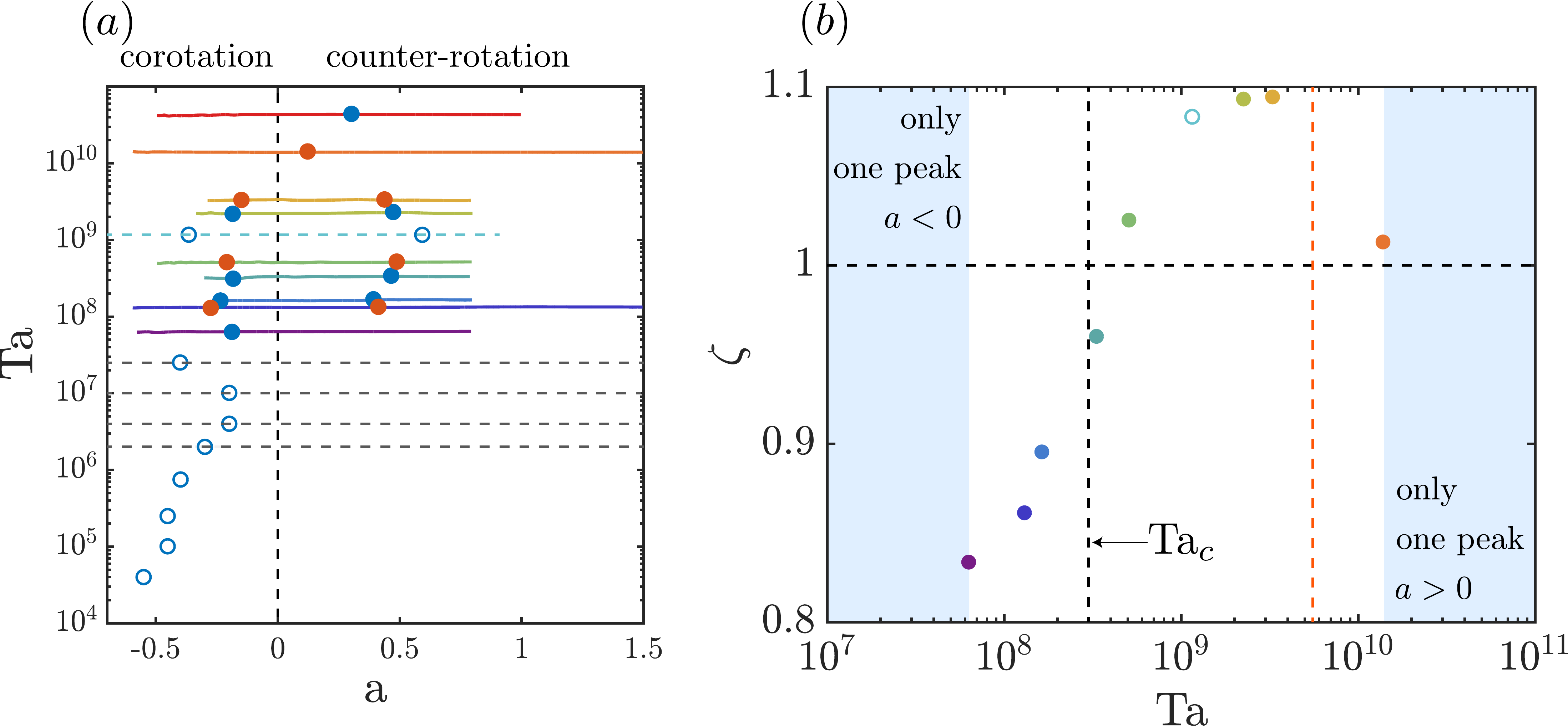}
 \caption{(a) Location in the phase space of the local maxima of angular momentum transport. The blue open circles for $\text{Ta} \leq 2.5\times 10^7$ are the DNS of \citet{ostilla2014}. The blue open circles located at $\text{Ta}=1.17\times 10^9$ are from DNS of the current study. The solid circles represent experimental data for the local maxima of angular momentum. The solid orange circles represent turbulent states where we perform PIV experiments as described in \sref{sec:piv} and shown in Figure \ref{fig:piv_points}. (b) Ratios of the magnitude of the angular momentum transport peaks as defined in eq.~\ref{eq:defzeta}. The colored points represent the experimental data. The open circle represents the DNS simulation shown in (a) for $\text{Ta}\approx 10^9$. The blue shaded areas represent turbulence levels wherein only one peak in angular momentum can be observed. The transition to the ultimate regime is observed at $\text{Ta} = \text{Ta}_\text{c} \approx 3 \times 10^8$. The vertical red dashed line represents the prediction of \cite{hannes2017} for the disappearance of the (broad) peak found in corotation, namely at $\Ta>4.95\times 10^9$. }
 \label{fig:maxima}
 \end{center}
 \end{figure}

In \fref{fig:maxima}a, we show the location of the observed local maxima throughout the parameter space ($a$, $\Ta$). Here, we also include the DNS data of \citet{ostilla2014} for the same radius ratio $\eta=0.91$ albeit for much lower values of Ta. We note that as the driving increases from $\Ta=\mathcal{O}(10^4)$ towards the critical Taylor number $\Ta_c$, the peak for corotation moves around, at times towards $a=0$, at others away from it. Past the transition, the location of this peak remains relatively stable at $a\approx -0.2 $ until it vanishes. Regarding the peak for counter-rotation, we see that it only appears when $\Ta>10^8$ and moves as the driving increases. When $1.33\times 10^8 \leq \Ta \leq 1.17\times 10^9$, the peak moves towards higher $a$ values of counter-rotation. However, for $\Ta>\mathcal{O}(10^{10})$, when only one peak is detectable, it seems to move back towards $a=0$. This side-effect of the disappearance of the broad peak means that the explanation by \citet{hannes2017} can be extended. We note, however, that at this driving, $\Nuw \Ta ^{-1/3}$ becomes less $a$-dependent which could over- or underestimate the precise location of the maximum. However, the shifting of the narrow peak is consistent with the asymptotic value of $a_{opt}$ at large $\Ta$ from \citet{ostilla2014} and happens around the same $\Ta$ for which the local effective exponent changes appeared in the $\Nuw(\Ta)$ relation. The reasons for this behaviour will be revisited in $\S$\ref{sec:merging}.
 
Interestingly, the value of the driving ($\Ta=5.1\times 10^8$) for which we detect two local maxima is close to the expected value of the transition to the ultimate regime, i.e. $\Ta_c=3\times 10^8$, and that at this $\Ta$, the relative magnitude of the peaks is very similar. In order to quantify this observation, we define the ratio of the magnitude of the peaks as

\begin{equation}
\zeta\equiv \dfrac{\Nuw(a=a_{\text{counter}})}{\Nuw(a=a_{\text{co}})}, \label{eq:defzeta}
\end{equation}

\noindent where $a_\text{co}$ and $a_\text{counter}$ denote the $a$-value that corresponds to the peak for co- and counter-rotation, respectively. In \fref{fig:maxima}b, we report $\zeta$ as a function of $\Ta$ showing that for $\Ta < \Ta_c$ we have $\zeta<1$, whereas for $\Ta > \Ta_c$ it holds $\zeta>1$. This yields an alternative representation of what was originally shown in \fref{fig:piv_points}: with sufficient driving the peak for counter-rotation will surpass the peak for corotation. As mentioned previously, the magnitude of both peaks seems to be nearly the same ($\zeta\approx 1$) close to the transition to the ultimate regime. This indicates the link between the appearance of the second peak, and the transitions of the boundary layer as postulated by \citet{hannes2017}. Furthermore, around the same values of $\Ta$, the dependence of $\Nuw$ on the roll wavelength, and thus on $\Gamma$ changes. This was seen as a crossing of the $\Nuw(\Ta)$ curves for different values of $\Gamma$ around the transition to the ultimate regime \citep{mar14,ostilla2014b}. That these phenomena occur all at the same time indicates the complex character of the transition to the ultimate regime. 

Finally, we note that for sufficiently large driving ($\Ta>4.95\times 10^{9}$), the narrow peak completely dominates and the broad peak can not be detected, as was also postulated by \citet{hannes2017}. At these values of $\Ta$, \cite{ostilla2014b} showed that the torque would no longer depend on roll wavelength, indicating again that the changes in the peak behavior are intimately linked to changes in the $\Nuw(\Gamma)$ relationships. 

\subsection{Local flow structure and its relation to the local $\text{Nu}_\omega$ maxima}
\label{sec:piv}

In the previous section, we showed that the narrow peak (counter-rotation) will surpass the broad peak (corotation) for values of driving $\Ta>4.95\times 10^{9}$. To further elucidate the mechanisms behind this phenomenon, we investigate the flow locally with PIV measurements. We explore a range of $\Ta$ which spans values before, close to and beyond the transition to the ultimate regime. For every driving, we investigate two flow states, namely where both the narrow and broad peaks are located as is shown in \fref{fig:piv_points}b and \tref{tab:table2}.

\begin{figure}
 \begin{center}
 \includegraphics[scale=0.5]{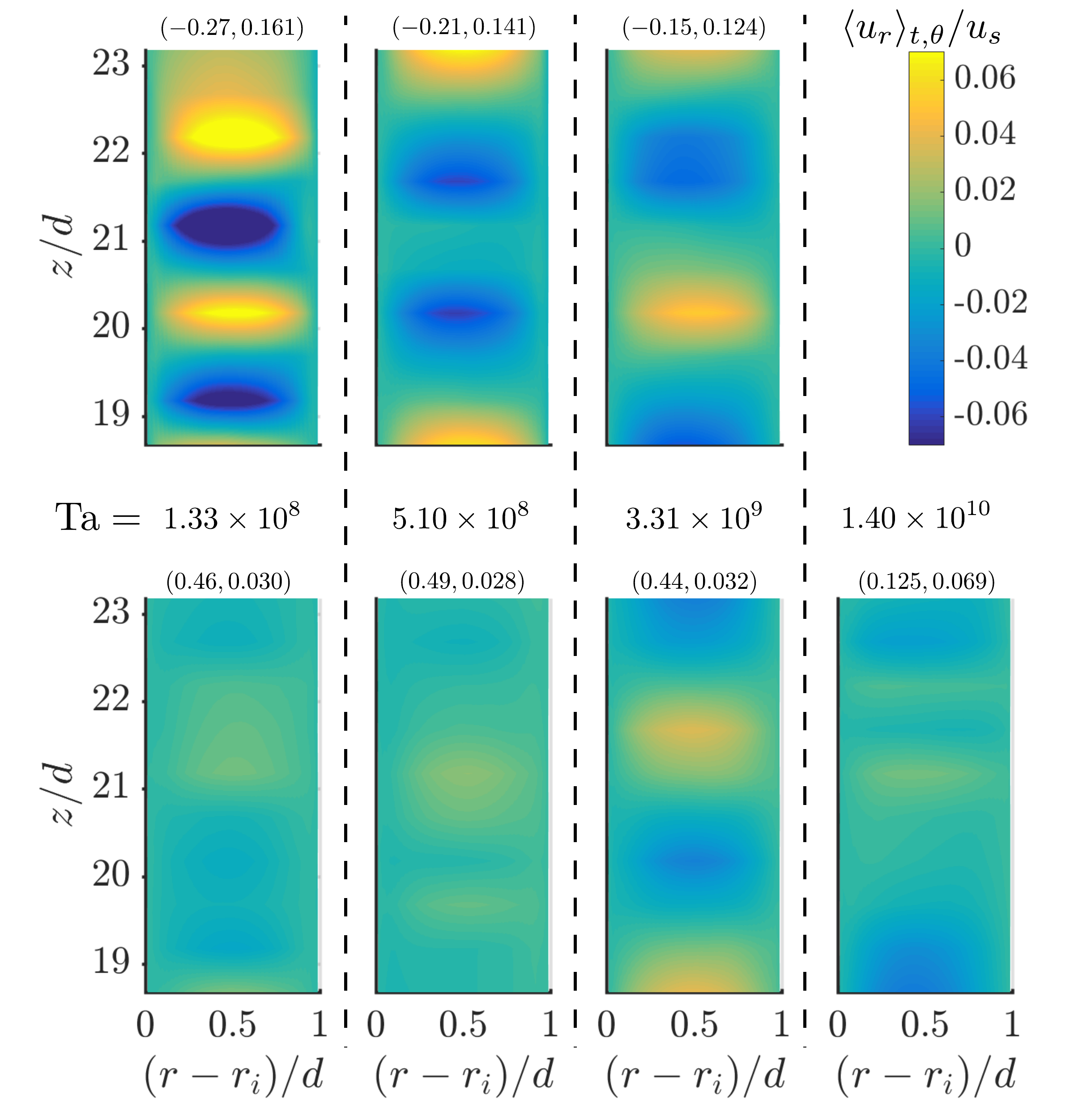}
 \caption{Azimuthally and time averaged normalized radial velocity obtained from PIV experiments as described in \fref{fig:piv_points}. The legend on top of each figure represents the value of ($a$,$R_\Omega$). The upper row represents measurements of the peak in the corotating regime while the bottom row shows measurements of the peak for counter-rotation. Along a single column, the Ta is fixed for both co- and counter rotating states. The dashed lines are added to emphasize the difference between the Ta values.}
 \label{fig:ur_exp}
 \end{center}
 \end{figure}

We first investigate the strength of the radial flow by looking at the time-azimuthally-averaged radial velocity, i.e.~$\langle u_r \rangle_{t,\theta}$. We remind the reader that every velocity field is measured in the $r-\theta$ plane, i.e.~$u_r=u_r(r,\theta,t)$. After an average over time and streamwise direction is performed, $u_r$ is only a function of the radius. We repeat this operation for all the heights explored which yields finally $u_r=u_r(r,z)$. In \fref{fig:ur_exp}, we show $u_r(r,z)$ for both peaks, which are located in the corotating and counter-rotating regime, respectively, and as function of $\Ta$. Here, we can clearly identify regions of negative and positive radial velocity along the axial direction which indicates the presence of Taylor rolls. Strikingly, we find that for all Ta explored, $|u_r|$ is much larger for corotation than for counter-rotation. This confirms what was shown numerically by \citet{hannes2016}: the broad peak (located for corotation) is accompanied by strong and coherent rolls.

In order to give a more quantitative picture of the strength of the rolls as a function of the driving, we look at the quantity

\begin{equation}
    \text{RMS}(\tilde{u_r})\equiv  \sqrt{\langle (\langle u_r /u_S, \rangle_{t,\theta,r_{\text{bulk}}})^2 \rangle_{z_\lambda}}.
\end{equation}

Here, $r_{\text{bulk}}$ denotes an average over the radial domain that defines the bulk region of the flow, namely $0.3 \leq (r-r_i)/d \leq 0.7$. Due to the presence of the vortical structures in the flow, we average over a distance $z_\lambda$ that corresponds to a roll wavelength. In order to find $z_\lambda$, we plot $\langle u_r \rangle_{t,\theta,r_{\text{bulk}}}$ as a function of $z$ (not shown). We define $z_\lambda$ as the distance that separates two adjacent maximum values of $\langle u_r \rangle_{t,\theta,r_{\text{bulk}}}$. In cases where there are no visible structures present---for example $a=0.46$, $a=0.49$, $a=0.125$---we simply take the average over $\Delta_z$. For the the case of $a=-0.21$, we also average over $\Delta_z$. The reason is that the maximum points of $\langle u_r \rangle_{t,\theta,r_{\text{bulk}}}$ in this case, are located at the minimum and maximum heights explored respectively.

\begin{figure}
 \begin{center}
 \includegraphics[width=\textwidth]{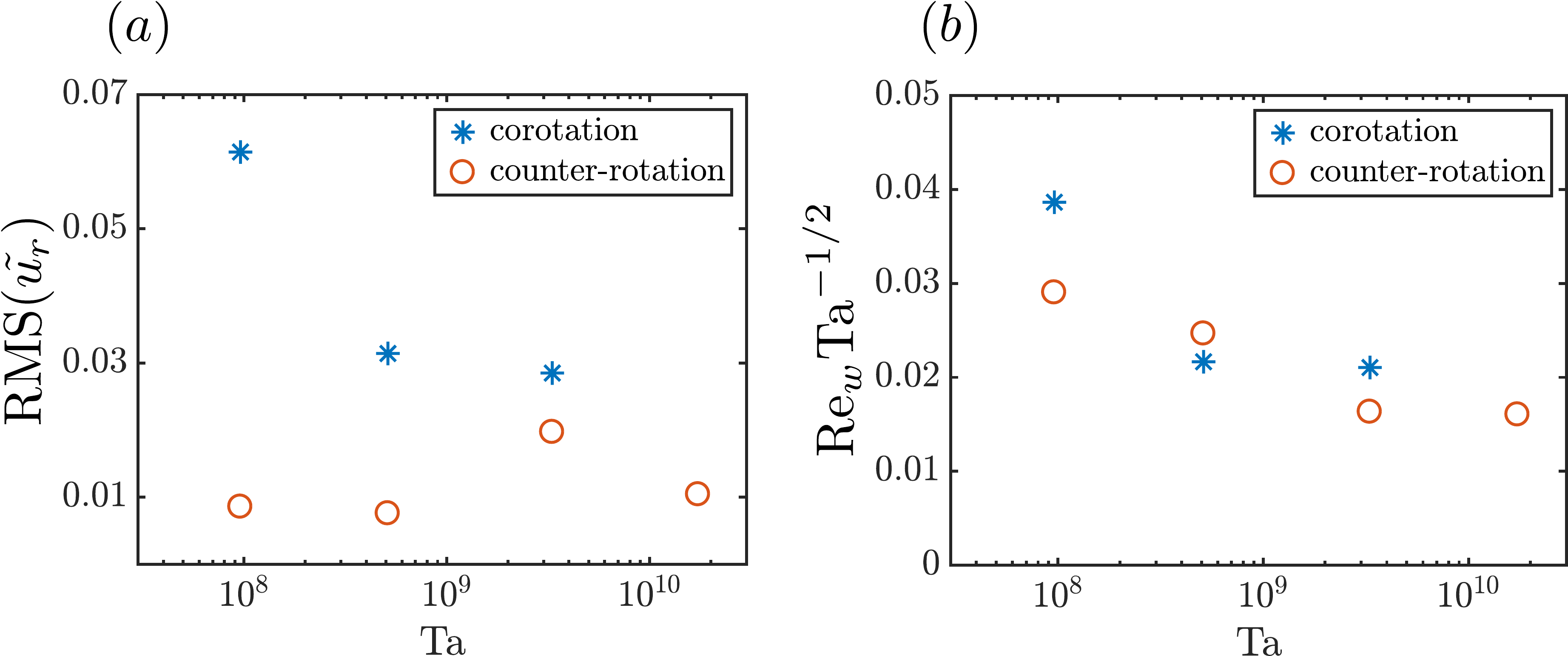}
 \caption{(a) RMS averaged along the axial direction of the azimuthally-time-bulk averaged radial velocity, normalized with the shear velocity $u_S$, as a function of the driving strength $\Ta$. (b) Wind Reynolds number (based on the radial velocity) as a function of $\Ta$. In both figures, the blue stars represent states that correspond to the peak in the corotating regime, while the red open circles represent measurements for the peak for counter-rotation.}
 \label{fig:wind}
 \end{center}
 \end{figure}

In \fref{fig:wind}a, we show $\text{RMS}(\tilde{u_r})$ as a function of the driving, where we observe that the $\text{RMS}(\tilde{u_r})$ of the co-rotating peak decreases with driving. The counter-rotating peak however, has a non-monotonic behaviour as a function of $\Ta$.

In addition to the strength of the rolls, we look now at the so-called ``wind'' by looking at the magnitude of the radial velocity fluctuations defined by

\begin{equation}
\sigma_{\text{bulk}}(u_r)\equiv \langle \sigma_{t,\theta}(u_r) \rangle_{r_{\text{bulk}},z_\lambda},
\end{equation}

\noindent where $\sigma_{t,\theta}(u_r)$ is the standard deviation profile of the azimuthal velocity and depends (for a fixed height) only on $r$. Here, the brackets $\langle \cdot \rangle$ denote the same average as the one performed to $\text{RMS}(\tilde{u_r})$. From this characteristic velocity we construct the wind Reynolds number, i.e. $\text{Re}_w=(d \sigma_{\text{bulk}}(u_r))/ \nu$. This is called the ``wind'' Reynolds number because, in analogy to heat transport in Rayleigh-B\'{e}nard (RB) flow, it is a measure of the energy of the flow that transports the conserved quantity \citep{grossmann2011}. So for the case of Taylor-Couette flow the strength of the wind is given by the (standard deviation of the) radial or axial velocity\citep{Huisman2012}.

In the classical regime of turbulence, the unifying theory of \citet{grossmann2011} predicts a scaling of the Reynolds wind $\text{Re}_w\propto \Ta^{3/7}$. When the driving is increased, towards the ultimate regime, the logarithmic corrections remarkably cancel out and an effective scaling of $\text{Re}_w\propto \Ta^{1/2}$ is observed \citep{grossmann2011,Huisman2012}. In \fref{fig:wind}b, we plot the compensated Reynolds wind with the scaling of the ultimate regime  $\textrm{Re}_w \Ta ^{-1/2}$ as a function of Ta. Here we see that, indeed, beyond the transition to the ultimate regime, $\text{Re}_w$ slowly asymptotes to a $\text{Ta}^{-1/2}$ scaling for both peaks, which is consistent with the observation of \cite{Huisman2012} for $\eta = 0.716$ at $a=0$.

We now draw the attention to the case of $\Ta=5.10 \times 10^8$, which is slightly above $\Ta_c$, and where both angular momentum peaks have roughly the same magnitude (see \fref{fig:maxima}b). As shown in \fref{fig:ur_exp}, for corotation at $a=-0.21$ (second panel of first row in \fref{fig:ur_exp}), a peculiar pattern of the radial velocity can be appreciated along the axial direction. Due to the shape and flow direction of rolls, we expect to see a succession of positive and negative radial velocities, as is seen in the first and third panel. Instead, in the second panel ($a=-0.21$), we encounter two consecutive regions of negative velocity, located in between $0.42 \leq z/L \leq 0.48$. This unexplained observation will be revisited in \sref{sec:roll_splitting}, as we unveil more data from the numerical simulations.

\section{Boundary layer transitions and state switching} 
\label{sec:simulations}

In the previous sections, we highlighted various observations which cannot be fully explained by the limited information we can retain from the experiments. In this section, we therefore turn to direct numerical simulations instead, such as to provide a more quantitative description of the observations. Namely, the mechanism responsible for the disappearance of the broad peak with sufficient shear (see \fref{fig:maxima}a), the observation of discrete jumps in the corotating regime ($a<0$) for $\Ta=2.23\times 10^9$ and $\Ta=3.31\times 10^9$ (see \fref{fig:piv_points}a) and the ``peculiar'' pattern of radial velocities which was presented in the previous section (see \fref{fig:ur_exp}). We closely examine the velocities, especially in the boundary layers, and inspect the changes in roll states.

\subsection{Disappearance of the broad peak}
\label{sec:merging}

\begin{figure}
\begin{center}
\includegraphics[scale=0.4]{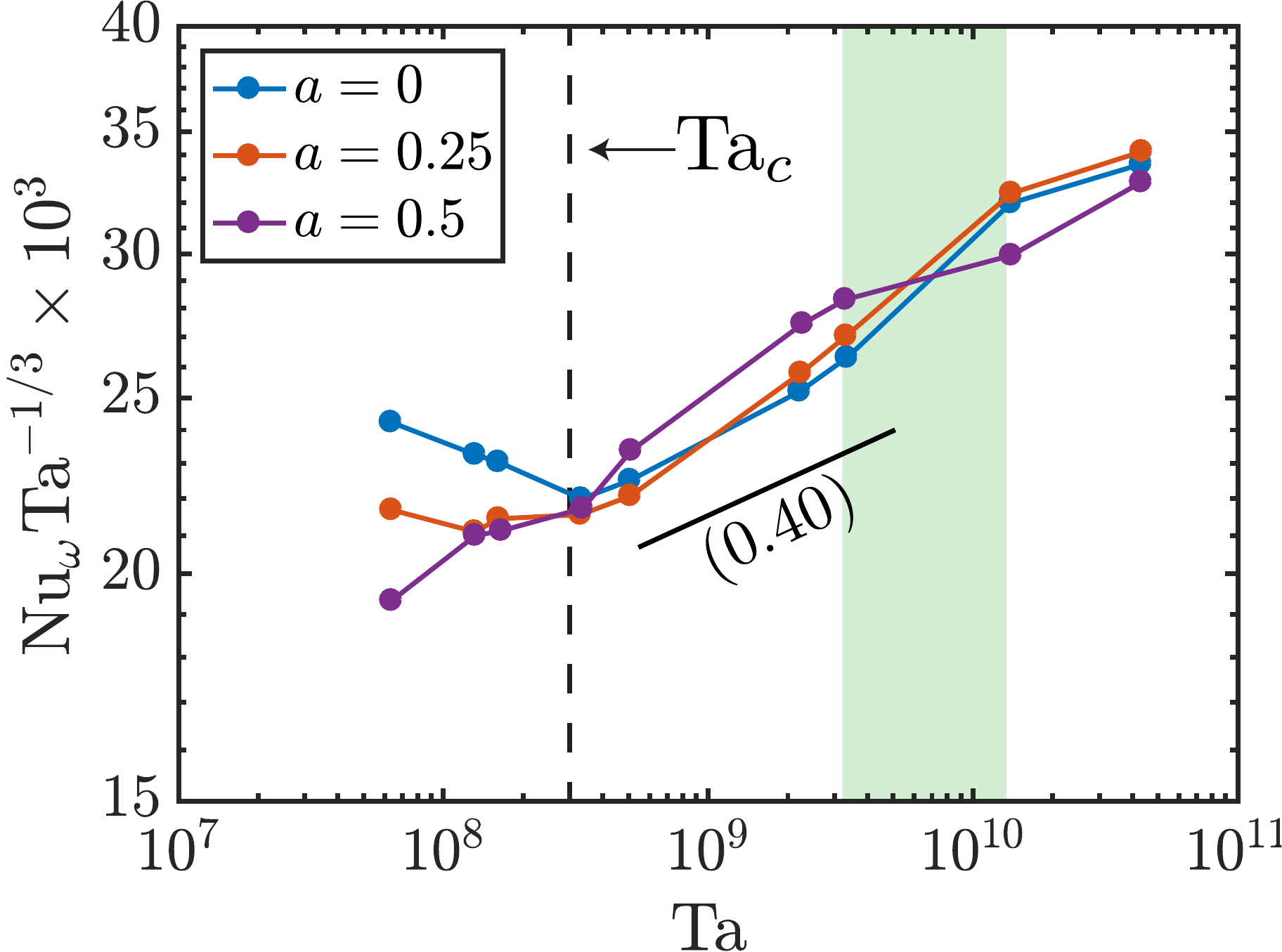}
\caption{Compensated Nusselt number as a function of $\Ta$ for three rotation ratios. A reordering of the curves around $\Ta\approx 10^{10}$ can be seen. The green shaded area corresponds to the region where a significant change in the local scaling exponent can be seen due to the disappearance of quiescent wind shearing regions. The black solid line representing the scaling of $\Nuw \propto \Ta^{0.4}$ is given as reference.}
\label{fig:multiple_a}
\end{center}
\end{figure}

We first focus on the changing shape of the $\Nuw(a)$ relationship at $\Ta\approx 10^{10}$ shown in  \fref{fig:piv_points}a. In section \ref{subsec:torque_measurements}, we mentioned how, for $a=0$, the sharp changes in the local scaling exponent of the $\Nuw(\Ta)$ relationship appeared at $\Ta\approx10^{10}$ (see \fref{fig:a_0}). These coincided with the disappearance of the torque on the roll-wavelength, and cannot be associated to the transition to the ultimate regime, as it happened at a much higher value of $\Ta$. In figure \ref{fig:multiple_a}, we show the behaviour of the $\Nuw(\Ta)$ curves for three selected values of $a$. We see how at $\Ta\approx 10^{10}$, a reordering of the curves occurs, distinct from the one seen at $\Ta=\Ta_c$. For $a=0$ and $a=0.25$, the changes in the local scaling exponents increase the effective exponent, while for $a=0.5$ the opposite process seems to occur.

\begin{figure}
\begin{center}
\includegraphics[width=\textwidth]{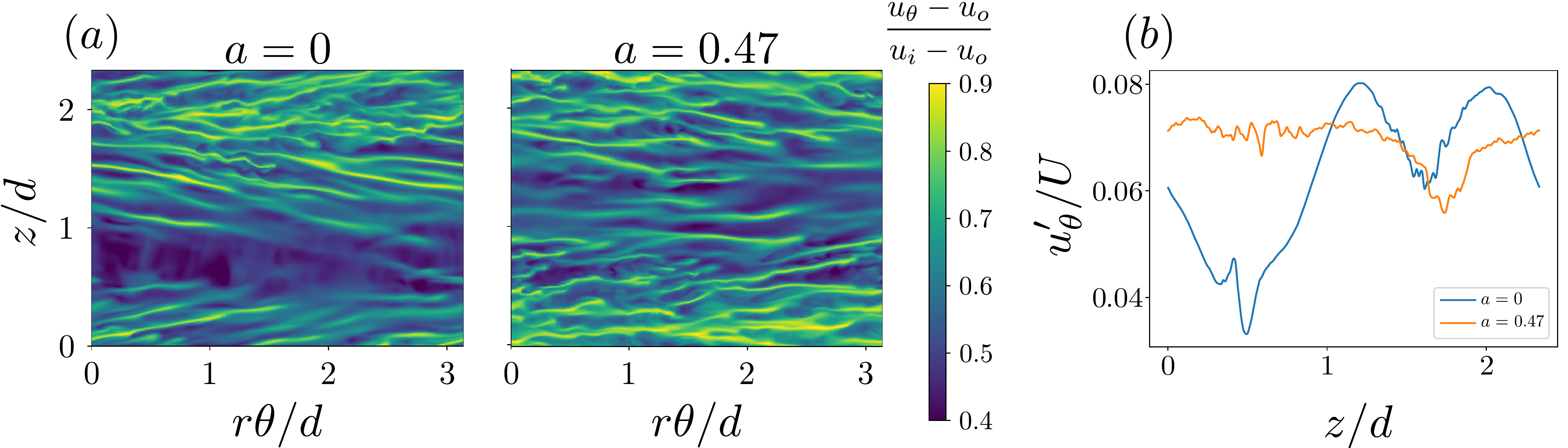}
\caption{(a) Instantaneous normalized azimuthal velocity at $r^+=15$ for $\Ta=5\times 10^8$ for $a=0$ ($R_\Omega=0.09$) and $a=0.47$ ($R_\Omega=0.03$). The DNS corresponds to the case of $\Gamma=2.33$. (b) Azimuthally average of the root-mean-squared azimuthal velocity $u_\theta^{'}$ as a function of axial position at $r^+=15$ for the two values of rotation $a$. A clear break in axial homogeneity can be appreciated for $a=0$.}
\label{fig:blsq1}
\end{center}
\end{figure}

By relating this to the transition observed in \citet{ostilla2014b}, where the quiescent regions of the boundary layer disappeared, we can explain why the counter-rotating maximum shifts. In figure \ref{fig:blsq1}, we show the structure of the near-wall region for several values of $R_\Omega$ (and $a$). As $R_\Omega$ is increased (i.e. from counter-rotation towards pure inner cylinder rotation), the flow structure self-organizes: the near-wall turbulent streaks occur in a stratified manner and the turbulent Taylor-roll is stabilized. This can be seen not only from the visual images of \fref{fig:blsq1}a, but also from the root-mean-squared values of $u^\prime_\theta$ shown in \fref{fig:blsq1}b. For $a=0$ they show a significant decrease in the regions which do not generate as many streaks, while there is a much smaller variation for $a=0.47$. This stratification of regions streaks can be linked to the ``plume-emission'' regions discussed in \citet{ost14}, where the regions with streaks were classified as plume-emitting, and the regions with no streaks were classified as plume-impacting.

The appearance of quiescent regions is not inconsistent with the idea that the creation of the narrow peak is due to shear instabilities that arise from the BLs. For our value of $\eta$, both before and after the transition to the ultimate regime, only parts of the boundary layer are active in producing plumes or streaks. During the transition to the ultimate regime, the plume--emitting part of the laminar boundary layer transitions to turbulence, while the quiescent part remains quiescent. For larger $\Ta$, beyond $\Ta=10^{10}$, the quiescent regions disappear, and the entire boundary layer becomes active and turbulent. As this happens, the $a=0$ curve surpasses the $a=0.5$ curve. Because there is no quiescent area to eliminate for $a>0.5$, these branches of the $\Nuw(\Ta)$ curve do not show significant change in the local scaling exponents and remain at a lower value. We note that this is unlike what occurs at $\eta=0.714$ in \citet{ost14}, where the disappearance of quiescent regions and the transition to the ultimate regime happens simultaneously.

\subsection{Roll state switches}
\label{sec:dns}
 
We now focus on the discrete jumps in the $\Nuw(a)$ curve at $\Ta = 2.23\times 10^9$ and $\Ta=3.31\times 10^9$ for $a<0$ from the experiments (see \fref{fig:piv_points}). We note that in previous simulations, both by \cite{hannes2017} and \cite{ostilla2014}, a small $\Gamma$ domain is used, which essentially fixes the roll size. The fact that no discrete jumps are detected in the numerics with small boxes, indicates that roll state switching might be responsible for this. Thus, in order to explain these jumps, we need to use both a small computational box which accommodates a single roll pair, as well as medium-sized boxes which can sustain changes in roll number. In this way, we can capture how the rolls manifest as the system goes from pure counter-rotation to corotation, and whether state switching takes place or not. We note that the number of rolls does not significantly affect the value of $\Nuw$ and that the only dependence comes from the roll wavelength \citep{bra13,ost16vlb}. However, by enforcing a periodicity length, both effects will manifest simultaneously. The roll number $N$ and the (non-dimensional) roll wavelength $\lambda_z$ are related by $\lambda_z=\Gamma/N$.

\begin{figure}
\begin{center}
\includegraphics[scale=0.4]{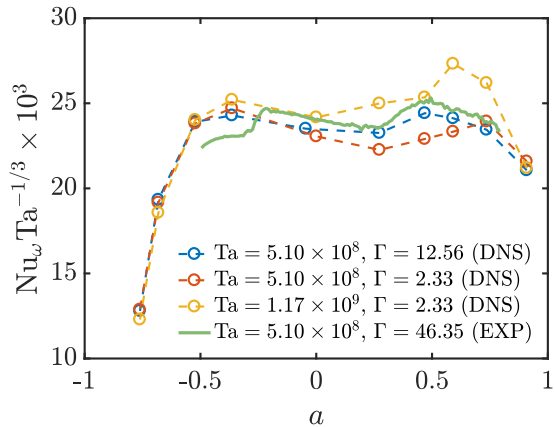}
\caption{Compensated Nusselt number as a function of $a$ obtained from the DNS of the current study. The green solid line is the experimental data performed at $\Ta=5.10\times 10^8$.}
\label{fig:nusselt_dns}
\end{center}
\end{figure}

In figure \ref{fig:nusselt_dns}, we show the compensated Nusselt number for both the small and the medium boxes for two $\Ta$, and the experiments for $\Ta=5.10\times 10^8$. The numerical simulations at $\Ta=5.10\times10^8$ reveal that for $a<-0.5$ the value of $\Nuw$ is at most weakly dependent on $\Gamma$. However, for $a\geq -0.5$ the two numerical studies and the experimental measurements result in different values of $\Nuw$. We note that $\Nuw$ was shown to have the largest dependence on the roll wavelength, while first and second order statistics also being affected in a small manner by \cite{ost16vlb}.

\begin{figure}
 \begin{center}
 \includegraphics[width=\textwidth]{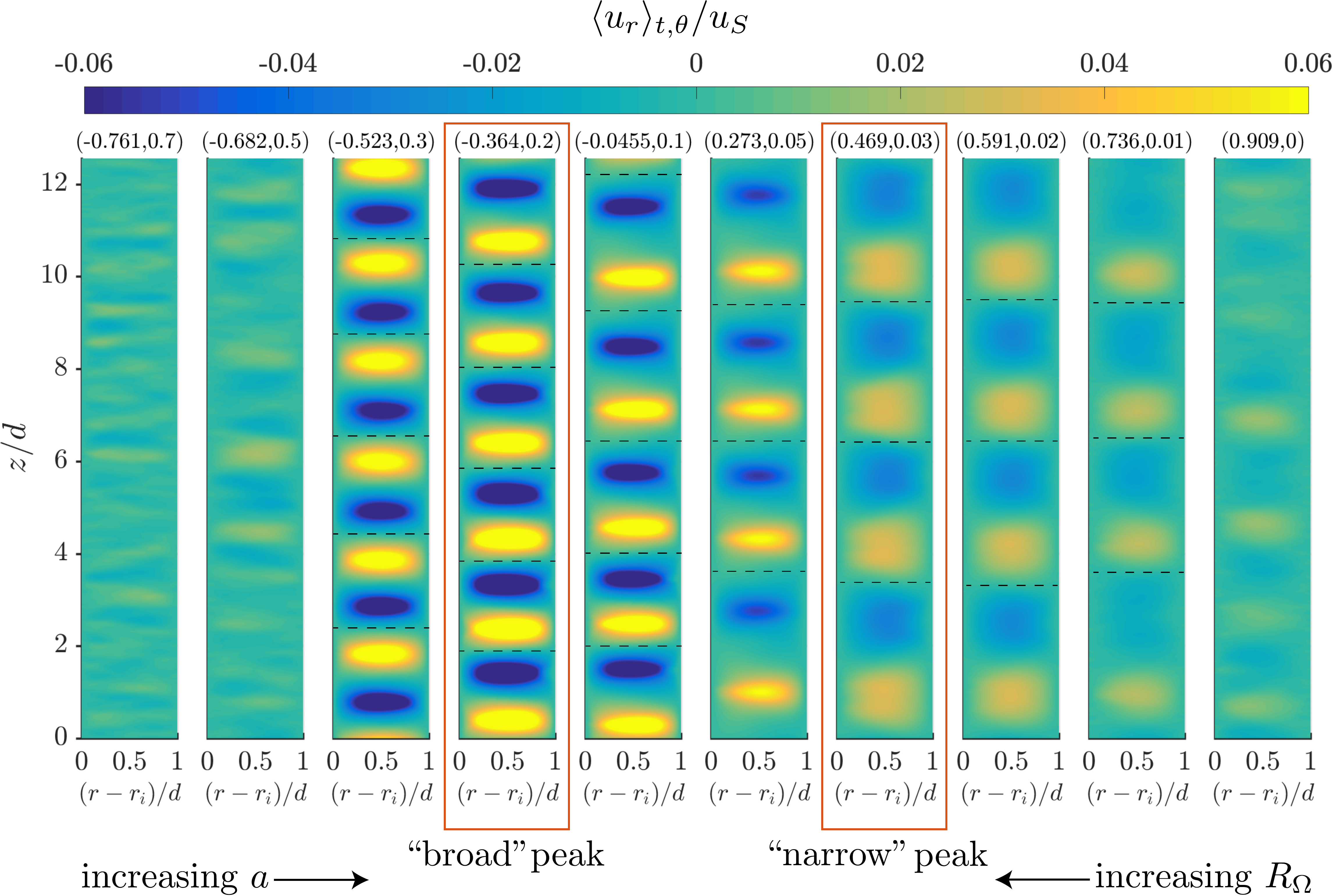}
 \caption{Azimuthally-time averaged radial velocity normalized with the characteristic velocity $U$ obtained from DNS for $\text{Ta}=5.10\times 10^8$ and $\Gamma=12.56$. Each plot represents a different rotation state, quantified by either $a$ or $R_\Omega$. The legend on top of each figure represents the value of ($a$,$R_\Omega$). The flow state that corresponds to either the ``broad'' or ``narrow'' peak are highlighted in red. The arrows indicate the direction of the increment of either $a$ or $R_\Omega$.}
 \label{fig:ur_dns}
 \end{center}
 \end{figure}

To further explore this, in figure \ref{fig:ur_dns} we show the azimuthally and temporally averaged radial velocity for the medium-sized domain ($\Gamma=12.56$), as we vary $a$ from the corotating to the counter-rotating regime. In terms of $R_\Omega$ this is equivalent to varying the Coriolis force from anti-cyclonic to zero. Here, we can see that the flow self-organizes in Taylor rolls as the Coriolis force starts to become dominant, i.e., when $R_\Omega \neq 0$ \citep{sac18}. As $R_\Omega$ increases, we first observe that the rolls become sharper and more prominent, as evidenced by an increment in $ |u_r|$, which is also observed in the experiments (see \fref{fig:ur_exp}). As we approach the ``broad peak'', the number of roll pairs switch, first from four to five, then from five to six, which sharply reduces the roll wavelength. This reduction in wavelength clearly has an effect on the torque as can be seen in \fref{fig:nusselt_dns} for $a>-0.5$  The change in the roll wavelength changes the proportion between quiescent and active regions inside the boundary layers  \citep{vanderpoel2011,vanderpoel2012}. As $\Ta$ increases, these sharp changes in the $\Nuw(a)$ curve disappear, because roll state switching does not modify the fraction of boundary layer regions which emit plumes.

The influence of the large structures on the torque can also be appreciated when we compare the $\Gamma=12.56$ simulation with the experimental data ($\Gamma=46.35$) shown in \fref{fig:nusselt_dns}. Here, we see that both the experimental and numerical data coincide within the region $0.3<a<1$, where the Coriolis force is small and the rolls only start to become organized. However, in the region between $-0.5<a<0.3$ there are significant discrepancies between experiments and numerics. These are probably a consequence of state switching. In the medium size box ($\Gamma=12.56$), a switch in the number of roll pairs can take place, but the values that the roll wavelength can take are more restricted due to the size domain and thus, leading to large variation in roll wavelength across different states. In the experiment $\Gamma$ is much larger, so state switches will generally have a very small effect on the roll wavelength, and as a consequence a small effect on the torque. This can explain both the jumps in the $\Nuw(a)$ curve, and why they are much smaller than for numerics.

 \subsection{Transient roll dynamics}
 \label{sec:roll_splitting}
 
During the transition to the ultimate regime, the size distribution of the roll changes (see \fref{fig:ur_dns}). This can be seen from the changing distance between the maximum values of $|u_r|$. This change in the roll dynamics suggests that there could be an interval of $R_\Omega$ in which the number of rolls is allowed to change, and it is connected to the pattern depicted at the end of \sref{sec:piv}, represented in the second panel of figure \ref{fig:ur_exp}. We can see there the presence of two similarly-signed radial velocities that are close to each other. If one would approach this from the point of view that Taylor rolls are present, it could only be possible to think that two rolls rotating in the same way are next to each other, as the region in between them is too small to allow the presence of a well formed counter-rotating roll. Otherwise, it could also be the case that this region has no rolls and we are encountering a local or transient phenomenon. In order to explore this event, we have to analyze the instantaneous velocity fields that DNS provides. Experimental results examine consecutive meridional planes ($r-\theta$) and assume that the flow is azimuthally homogeneous. However, if the structures change only locally, this could be wiped out by an averaging operation. 

As $R_\Omega$ increases (see \fref{fig:ur_dns}), the rolls become sharper, until at a certain point, one of them can start to split up locally, such that transiently, outflow, or inflow regions with the same sign coexist very close to each other at certain values of the azimuth. As $R_\Omega$ is further increased, the roll splits up completely to form a new roll pair. However, the speed at which this roll ``dislocation'' exists and propagates to ``fracture'' the roll is a priori unknown.

  \begin{figure}
 \begin{center}
 \includegraphics[width=\textwidth]{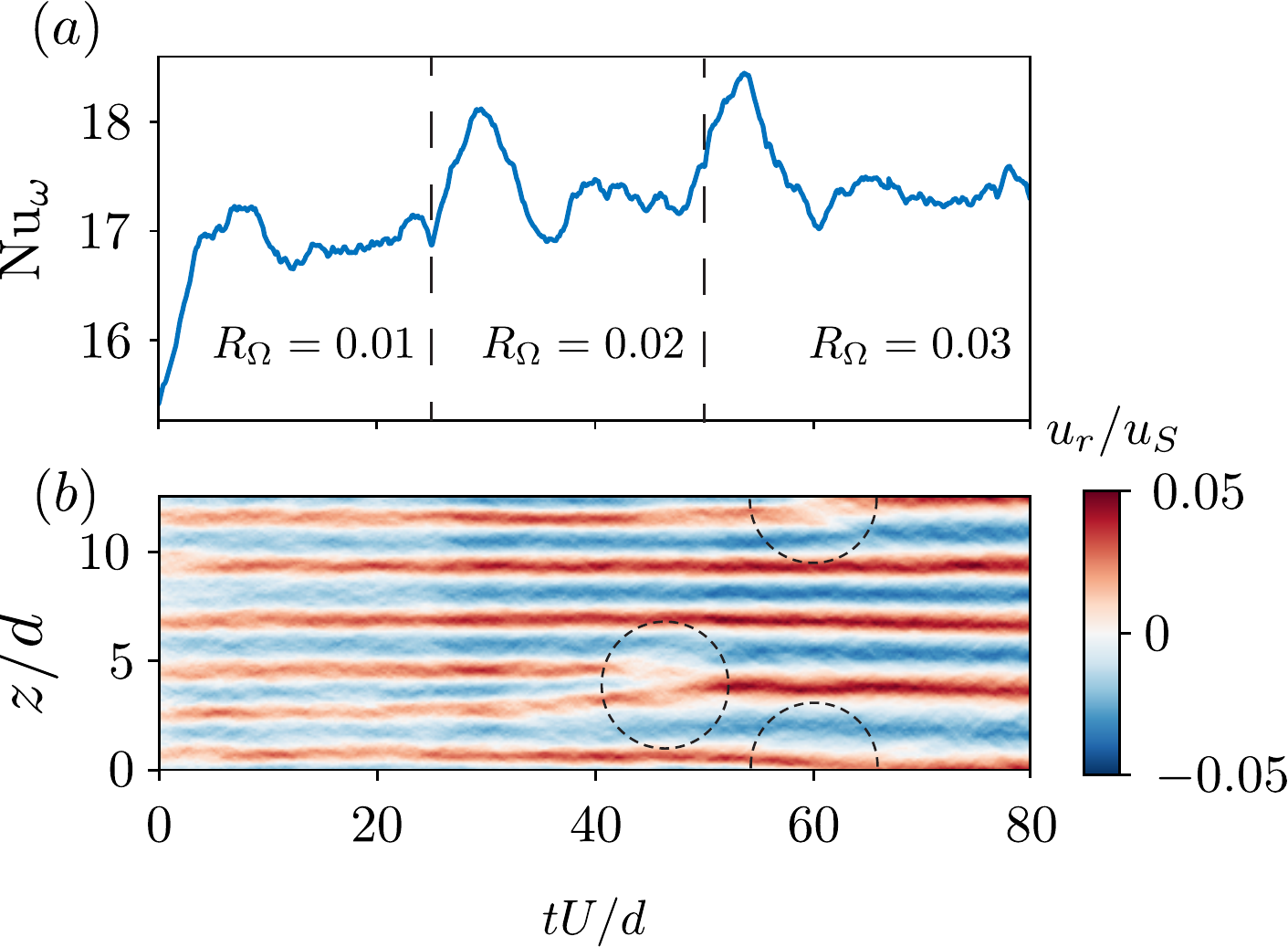}
 \caption{DNS results of varying $R_\Omega$ for $\text{Ta}=5.10\times 10^8$ and $\Gamma=12.56$. (a) Temporal evolution of the dissipation  $\Nuw$. (b) Time-space diagram of the instantaneous $u_r$ at mid-gap. Here, two regions of merging rolls can clearly be seen, which are generated from two consecutive outflow regions. These mergers are highlighted with the dashed circles The vertical dashed lines in (a) represent a stage during the simulation for a fixed value of $R_\Omega$.}
 \label{fig:rollsmerge}
 \end{center}
 \end{figure}

In order to closely inspect the change in the morphology of the rolls as a function of the Coriolis force, we perform DNS by slowly changing the value of $R_\Omega$ over a certain number of large eddy turnover times $\tau=tU/d$, where $U$ is the characteristic velocity as defined in section \ref{sec:num_details}. The simulations are performed for $\Ta=5.10\times 10^8$ and for the medium-sized box ($\Gamma=12.56$). The idea is to capture the evolution of the radial velocity and its effect on $\Nuw$ as we slowly increase $R_\Omega$ in discrete steps. We initiate the simulation at $\tau=0$ with a statistical stationary TC flow and $R_\Omega=0$. Next, at $\tau=25$, we increase the value of the Coriolis force to $R_\Omega=0.02$. When $\tau=50$ is reached, we set finally $R_\Omega$ to 0.03. In \fref{fig:rollsmerge}a, we show the instantaneous $\Nuw$ and in \fref{fig:rollsmerge}b, we show the space-time evolution of the instantaneous radial velocity $u_r$ at mid-gap. We note that, globally, a correlation between transient regions of $\Nuw$ and the change in $R_\Omega$ can be observed. This is caused by the increasing acceleration of the flow, in which the number of rolls changes as well. Locally, we clearly see two instants in which two consecutive rolls begin to approach each other until they merge. During this transient phenomenon, that lasts $\approx 10\tau$, the shape of the rolls changes, and at the end also the global number of rolls couple, due to the merging event. During this time two consecutive outflow regions are allowed to coexist close to each other, while the inflow region slowly disappear. This indicates---just as it was observed in the experiments shown in \fref{fig:ur_exp}---that close to the transition to the ultimate regime, transient events and changes in the roll dynamics are allowed under suitable conditions.

\section{Summary and conclusions}
\label{sec:conclusion}
We probe the angular momentum transport with both experiments and direct numerical simulations for $\eta=0.91$ as a function of the driving which we quantify with $\Ta$ and the magnitude of the Coriolis force which we quantify with either the rotation ratio $a$ or the rotation number $R_\Omega$. The range of shear driving we explore spans several decades of $\Ta$, namely $\mathcal{O}(10^7)-\mathcal{O}(10^{10})$, which includes the transition to the ultimate regime at $\Ta\approx \Ta_c=3\times 10^8$. In the vicinity of $\Ta_c$, our experiments reveal the presence of two local angular momentum transport maxima for fixed $\Ta$, as it was firstly reported by the numerical works of \citet{hannes2016} and \citet{hannes2017}. Thus we have demonstrated that this occurrence is not an artifact of the axial domain used in the numerics. We showed that the broad peak is associated to the strengthening of the large-scale structures (Taylor-rolls) due to an increment in centrifugal forcing $R_\Omega$, as it is evidenced by both PIV and DNS. In addition, the numerical simulations reveal that the appearance of the narrow peak is a consequence of shear instabilities that arise from the BLs. 
Moreover, we show that as the driving increases, the broad peak remains roughly at the same $R_\Omega$ while decreasing its magnitude. Conversely, the narrow peak increases its magnitude while experiencing a shift in $R_\Omega$. We attribute this to the different way in which the near-wall structures self-organizes at $\Ta\approx 10^{10}$. Explicitly, the disappearance of quiescent wind sheared regions at the cylinder walls, which suddenly transition to a state where the whole BLs can emit plumes, as evidenced by a sharp localized increase in the effective exponent of the $\Nuw(\Ta)$ relationship.
With sufficient driving, both the experiments and numerics confirm that the shear instabilities dominate over centrifugal instabilities and the broad peak disappears at around $\Ta=4.95 \times 10^9$. Beyond this value, only the narrow peak can be detected. 

In the experiments, a peculiar state of the Taylor rolls is observed at $a=-0.210$ ($R_\Omega=0.141$) and $\Ta=5.10\times 10^8 $, which is close to $\Ta_c$, where both the narrow and broad peaks are present and have roughly the same magnitude. Here, we observe two neighboring Taylor rolls which rotate opposite to each other. We explore the possibility that this \textit{unusual} state is produced by roll splitting, which occurs  when the axial extent is sufficiently large and for sufficient $R_\Omega$. The numerics at $\Ta=5.10\times 10^8$ reveal that as the magnitude of the Coriolis force increases via $R_\Omega$, the Taylor rolls undergo a reorganization and a change in both their number and wavelength. This effect translates into a change in the global torque, i.e. $\Nuw$, which is evidenced by different $\Nuw (a)$ curves for different values of $\Gamma$. We explore the reorganization of the rolls in detail by performing a last numerical simulation where we slowly vary $R_\Omega$ several times. Here, we observe that for the same driving $\Ta$ and curvature $\eta$ as in the experiments (different $\Gamma$), the merging of two adjacent rolls is possible. This suggests that the ``peculiar'' pattern of roll formation discussed previously can be a consequence of such long-time transient effects. Further work in this direction is encouraged in order to thoroughly explore this new feature of secondary flows in TC flow.

\section*{Acknowledgements}
This paper is dedicated to Bruno Eckhardt, who unexpectedly passed away in July 2019. Over the last $15$ years, we had various stimulating discussions with him on TC turbulence in general and over the last few years in particular on the double-peak nature of the angular momentum transport, for which we are really thankful. We acknowledge Gert-Wim Bruggert for technical support. This work was funded by STW, FOM, and MCEC, which are part of the Netherlands Organization for Scientific Research (NWO). Chao Sun acknowledges financial support the Natural Science Foundation of China under Grant No. 11988102, 91852202 and 11672156. We thank the Center for Advanced Computing and Data Science (CACDS) at the University of Houston for providing computing resources and we also acknowledge PRACE for awarding us access through the John von Neumann Institute for Computing (NIC) on the GCS Supercomputer JUWELS at J\"ulich Supercomputing Centre (JSC), under PRACE project number 2017174146. Declaration of interests: The authors report no conflict of interest.

\bibliographystyle{jfm}
\bibliography{bibliography}

\end{document}